\documentclass[fleqn,10pt]{wlscirep}
\usepackage[utf8]{inputenc}
\usepackage{amsfonts,amssymb,amsmath}
\usepackage{graphicx}

\usepackage{dcolumn}
\usepackage{bm}
\usepackage{xcolor}
\usepackage{subcaption}
\usepackage{siunitx} %\SI{100}{\micro\meter}
\usepackage{epsfig}
\usepackage{cleveref}
\usepackage{ulem}% pour barrer le texte

\graphicspath{{./figures/}}

\crefname{figure}{Figure}{Figures}
\crefname{equation}{Equation}{Equations}
\crefname{table}{Table}{Tables}

\DeclareSIUnit\pxl{pixels}
\DeclareSIUnit\ppm{ppm}
\DeclareSIUnit\ppb{ppb}
\DeclareSIUnit\ppt{ppt}
\DeclareSIUnit{\liter}{$\ell$}

\def \draft {1}

\usepackage{xparse}
\usepackage{ifthen}
\DeclareDocumentCommand{\comment}{m o o o o}
{\ifthenelse{\draft=1}{
    \textcolor{red}{\textbf{C : }#1}
    \IfValueT{#2}{\textcolor{blue}{\textbf{A1 : }#2}}
    \IfValueT{#3}{\textcolor{ForestGreen}{\textbf{A2 : }#3}}
    \IfValueT{#4}{\textcolor{red!50!blue}{\textbf{A3 : }#4}}
    \IfValueT{#5}{\textcolor{Aquamarine}{\textbf{A4 : }#5}}
 }{}
}

\newcommand{\todo}[1]{
\ifthenelse{\draft=1}{\textcolor{red!50!blue}{\textbf{TODO : \textit{#1}}}}{}
}

\title{Quantitative analysis of self-organized patterns in ombrotrophic peatlands}

\author[1,2]{Chloé Béguin}
\author[1,2]{Maura Brunetti}
\author[1,2,*]{Jérôme Kasparian}

\affil[1]{Group of Applied Physics, University of Geneva, Chemin de Pinchat 22, 1211 Geneva 4, Switzerland}
\affil[2]{Institute for Environmental Sciences, University of Geneva, bd Carl Vogt  66, 1211 Geneva 4, Switzerland}
\affil[*]{jerome.kasparian@unige.ch}

\begin{abstract}
We numerically investigate a diffusion-reaction model of an ombrotrophic peatland implementing a Turing instability relying on nutrient accumulation. We propose a systematic and quantitative sorting of the vegetation patterns, based on the statistical analysis of the numbers and filling factor of clusters of both \textit{Sphagnum} mosses and vascular plants. In particular, we define the transition from \textit{Sphagnum}-percolating to vascular plant-percolating patterns as the nutrient availability is increased.
Our pattern sorting allows us to characterize the peatland pattern stability under climate stress, including strong drought.
\end{abstract}

\begin{document}

\flushbottom
\maketitle

\thispagestyle{empty}

%%%%%%%%%%%%%%%%%%
\section*{Introduction}

Peatlands are humid ecosystems, poor in O$_2$ and often in nutrients, with over 65\% of organic contents over dry soil mass~\cite{Charman2002}. Covering around \SI{4000000}{km^2} worldwide, they constitute an essential carbon well storing \SIrange{270}{550}{PgC}~\cite{Quillet2013}, an amount comparable to that of carbon released into the atmosphere since 1750~\cite{Ciais2014}. Besides their contribution to the carbon cycle, the stratified accumulation of organic matter in a relatively preserved environment constitutes remarkable archives of past climate. 
Their dynamics is mainly governed by their hydrology~\cite{Charman2002,Thompson2017}. Ombrotrophic peatlands, that are isolated from surface and subsurface water flows,  primary rely on precipitations~\cite{Winter2000}.  

First attempts to model peatlands primarily aimed at describing peat formation, with a focus on the vertical dimension~\cite{Clymo1984,Hilbert2000,Frolking2010,Baird2012}. 
However, like many ecosystems~\cite{Rietkerk2004,Koppel2008}, peatlands display self-organized spatial patterning, that is generally considered stable although it may slowly evolve in time~\cite{Kettridge2012}. The peatland patterns are believed to be resilient against climate changes~\cite{Kettridge2014,Waddington2015,Schneider2016,Thompson2017}.
Based on two-dimensional modelling, several processes have been proposed to create or influence these patterns, including peat accumulation, water ponding, and nutrient accumulation~\cite{Eppinga2008,Baird2012,Morris2013}. In the following, we shall focus on the latter, based on numerical~\cite{Eppinga2008} and empirical~\cite{Eppinga2009b} arguments. Nutrient accumulation can be described as a reaction-diffusion process, in a two-dimensional model. The active transport of nutrients towards spots of vascular plant like \textit{Carex}, results in their accumulation at specific locations where the subsequent development of vascular plants is favored and in their depletion elsewhere~\cite{Rietkerk2004a}. 
This model was refined by considering the competition between vascular plants and \textit{Sphagnum} mosses~\cite{Eppinga2009a}, leading to sharper and more stable patterns.

Here, we analyze the biomass-nutrient interaction in terms of a Turing instability~\cite{Kukushkin1999,Rietkerk2008,Oppo2009}, relying on the strong discrepancy between the fast and slow diffusion of nutrients and vascular plant, respectively.
We further build on this class of two-dimensional reaction-diffusion model to systematically analyze and classify the resulting patterns, based on statistics of the number and filling factor of clusters of vascular plants and  \textit{Sphagnum}. We also investigate the equivalent phase transition from \textit{Sphagnum} and vascular plant cluster percolation, and characterize the associated scaling law.
Finally, we apply our classification to describe the peatland pattern stability especially under climate stress, including strong prolonged drought. 

%%%%%%%%%%%%%%%%%%
\section*{Peatland model description} \label{sec:modelDescription} 

\begin{figure}[tb]
\centering
\includegraphics[width=0.7\columnwidth, keepaspectratio]{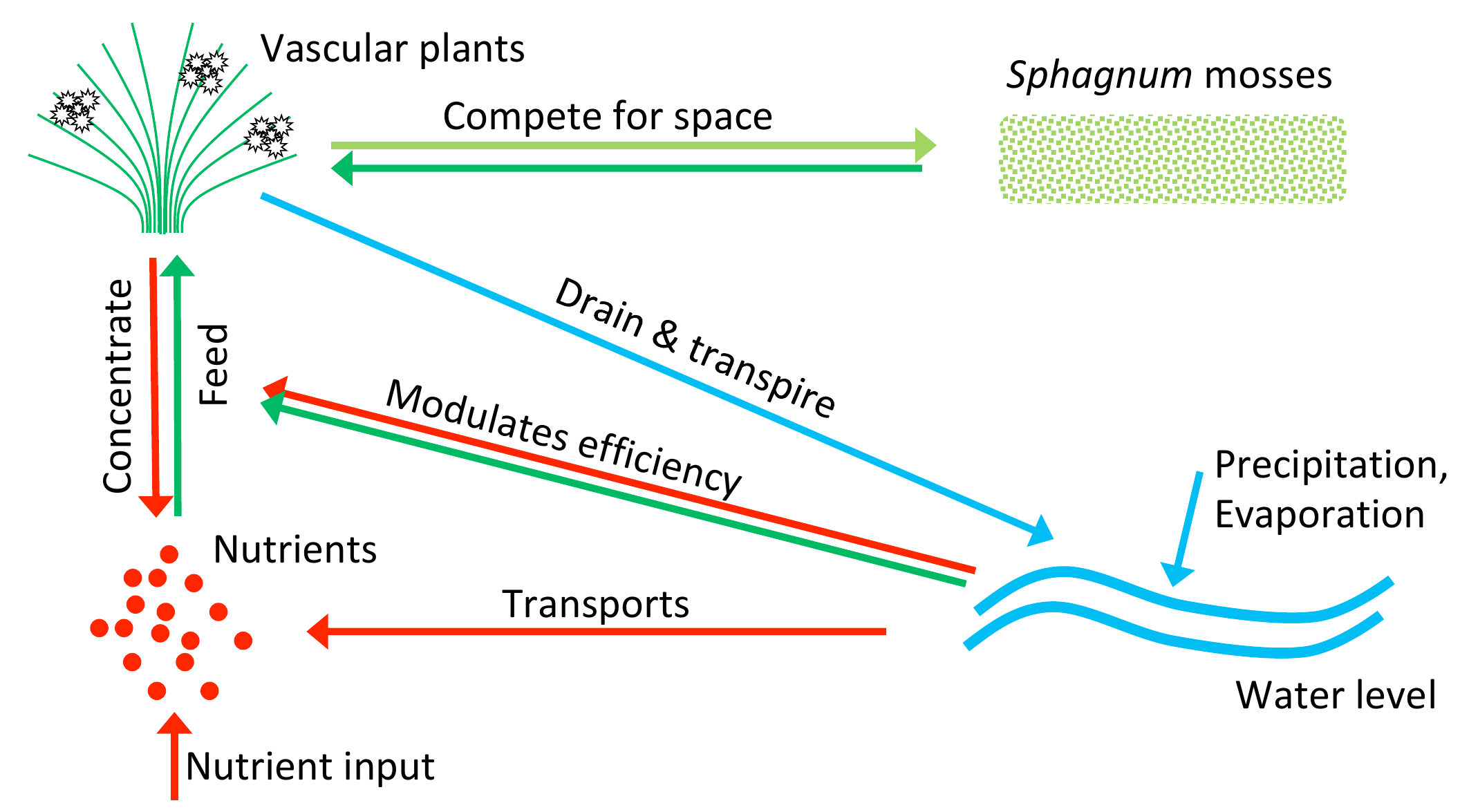}
\caption{Main interactions between the four sub-systems of the peatland model.}
\label{fig:interactions}
\end{figure}

Our work is based on Eppinga \textit{et al.}'s two-dimensional reaction-diffusion model~\cite{Eppinga2009a}, which describes the interactions between vascular plants, \textit{Sphagnum}, nutrient availability and hydraulic head (\Cref{fig:interactions}). The heart of the interaction is nutrient accumulation due to water draining by vascular plants, via transpiration. Vascular plants also promote water loss by their transpiration and compete for space with \textit{Sphagnum}.

More specifically, the model relies on four partial differential equations describing the temporal evolution of the following spatial variables: vascular plant and \textit{Sphagnum} mosses biomass per unit surface $B(x,y,t)$ and $S(x,y,t)$, hydraulic head $H(x,y,t)$, and nutrient availability $N(x,y,t)$, the latter being linked to the nutrient concentration per unit surface $N_\textrm{c}(x,y,t)$ (both in \si{g/m^2}) by 
\begin{equation}
N = N_\textrm{c} \ H \ \theta
\end{equation}
where $\theta$ describes the soil porosity.
The vascular plant biomass density evolves as:
\begin{equation}
\frac{\partial B}{\partial t} = r_\textrm{B} N_\textrm{c} B f(H) - (d+b) B - c_\textrm{BS}S - \frac{k_\textrm{BS}SB}{H_B + B}
	+ D_\textrm{B}  \left(\frac{\partial^2 B}{\partial x^2}+\frac{\partial^2B}{\partial y^2}\right)
\label{eq:B}
\end{equation}
where the first term is the growth of the plants, $r_\textrm{B}$ being a growth parameter and the water-stress function $f(H)$ defines water availability ($f(H) = 0$ if the water level is below the vascular plant root depth, 1 if above a threshold level $h_1+z$, and evolves linearly in-between). The second term encodes the vascular losses through death ($d$) and other causes ($b$). The third and fourth terms represent local environmental modifications due to the presence of \textit{Sphagnum} mosses, as detailed in~\cite{Eppinga2009a}. The last term accounts for diffusion. \Cref{table:parameters} summarises the meaning and default value of each variable. 

The growth of vascular plants is regulated by their access to nutrients. In an ombrotrophic peatland, the nutrient external input $N_\textrm{in}$ mostly originates from the atmosphere, via precipitation~\cite{Proctor1992,Charman2002}. The nutrient availability $N$ is then governed by losses through the uptake by the vascular plants (with $u$ the uptake parameter), the release of nutrients by dying vascular biomass, other losses characterised by the parameter $r$, and nutrient transport allowing the vascular plants to harvest nutrients from their neighbourhood. This transport occurs both by diffusion according to Fick's law and through advection with the groundwater. {The latter is described by the Darcy flow, which expresses fluid flows at Reynolds numbers below 1--10}, i.e. the regime where inertial effects are negligible, so that the flow occurs along the pressure gradient. This regime is typical of geophysical flows, including the diffusion of water through porous materials. Thus, the nutrient availability evolves as
\begin{equation}
\frac{\partial N}{\partial t} = N_\textrm{in} - u N_\textrm{c} B f(H) + \frac{d u}{g}B - rN 
	+ \theta D_\textrm{N} H \left(\frac{\partial^2 N_\textrm{c}}{\partial x^2}+\frac{\partial^2 N_\textrm{c}}{\partial y^2}\right) 
	+ k H \biggl[\frac{\partial}{\partial x}\left(N_\textrm{c}\frac{\partial H}{\partial x}\right) + \frac{\partial}{\partial y}\left(N_\textrm{c}\frac{\partial H}{\partial y}\right)\biggl]
	\label{eq:N}
\end{equation}
where $k$ is the hydraulic conductivity. The vascular plant access to nutrients is modulated by the access to water. The hydraulic head $H$ evolves under the influence of the precipitations $p$ (with soil porosity parameter $\theta$), the evaporation parameter $e$, and the transpiration parameter $t_\textrm{v}$:
\begin{equation}
\frac{\partial H}{\partial t} = \frac{p}{\theta} - \frac{e}{\theta}f(H) - \frac{{t_\textrm{v}}}{\theta}f(H)B
	 + \frac{k}{\theta}\biggl[\frac{\partial}{\partial x}\left(H\frac{\partial H}{\partial x}\right) + \frac{\partial}{\partial y}\left(H\frac{\partial H}{\partial y}\right)\biggl]
	 \label{eq:H}
\end{equation}
Note that the transpiration only depends on {the vascular plant biomass $B$}, as \textit{Sphagnum} mosses have no vascular system and do not transpire. The last term is the Darcy Flow.
Finally, vascular plants compete for space with \textit{Sphagnum} mosses, the biomass $S$ of which is given by
\begin{equation}
\frac{\partial S}{\partial t} = r_\textrm{s} S \left(1-\frac{S}{S_\textrm{max}}\right) - \frac{k_\textrm{SB}SB}{H_\textrm{S} + S} + D_\textrm{S}  \left(\frac{\partial^2 S}{\partial x^2}+\frac{\partial^2 S}{\partial y^2}\right) 
\label{eq:S}
\end{equation}
where the first term describes a logistic growth, while the second one corresponds to competition with vascular plants, and the last one is diffusion. All parameters driving the model are spatially homogeneous and constant. 
They are defined as mean annual values, without seasonal nor daily variability. The numerical implementation is detailed in the Methods section below.

It should also be pointed out that the model primarily aims at characterizing the self-organized patterns, but does not seek perfect realism. In particular, the model is restricted to two types of plants, namely mosses and a generic vascular plant, not distinguishing between, e.g., trees, shrubs, and sedges~\cite{Charman2002}. The evolution of the soil, including peat formation, decomposition, and loss~\cite{Tfaily2018} and the associated deviation of the peatland surface from a planar one is also disregarded, preventing the model from realistically describing the differential soil porosity and conductivity to water~\cite{Price1992,Eppinga2009,Morris2015}. Finally, the periodic boundary conditions and the flat topography exclude any net water flow through and into the simulation area~\cite{Winter2000}, and restrict water input to precipitations~\cite{Morris2013}.

\begin{table*}
\begin{tabular}{llll}
\hline
Symbol & Interpretation & Unit & Default value \\
\hline
\multicolumn{4}{l} {Variables} \\
\hline
$B$ & Vascular plant piomass & \si{g_B m^{-2}} & Variable \\
$S$ & \textit{Sphagnum} biomass & \si{g_S m^{-2}}  & Variable \\
$H$ & Hydraulic head & \si{m} & Variable \\
$N$ & Nutrient availability & \si{g_N m^{-2}} & Variable \\
$N_\textrm{c}$ & Nutrient concentration & \si{g_N m^{-3}} & Variable \\
\hline
\multicolumn{4}{l} {water-stress function: $f(H) = \dfrac{H-z-h_2}{h_1-h_2}$ for $h_1 \ge H-z \ge h_2$, $f(H\le h_2+z)=0$, $f(H\ge h_1+z)=1$}  \\
\hline
$h$ & Pressure head & \si{m} & Variable \\
$h_1$ & Pressure head below which soil water stress occurs & \si{m} & 0 \\
$h_2$ & Root depth & \si{m} & -0.5\\
$z$ & Elevation head & \si{m} & 1 \\
\hline
\multicolumn{4}{l} {Vascular biomass equation}\\
\hline
$r_\textrm{B}$ & Vascular plant growth parameter & \si{m^3 g_N^{-1} yr^{-1}} &  0.2\\
$d $ & Vascular plant recycling parameter & \si{yr^{-1}} & 0.1 \\
$b$ & Vascular plant loss parameter & \si{yr^{-1}} & 0.2 \\
$D_\textrm{B}$ & Vascular plant diffusion parameter & \si{m^2 yr^{-1}} & 2 \\
$c_\textrm{BS} $ &  \textit{Sphagnum} inhibiting coefficient on vascular biomass & \si{g_B g_S^{-1} yr^{-1}} & 0.1 \\
$k_\textrm{BS}$ & Feedback coefficient for vascular plants & \si{g_B g_S^{-1} yr^{-1}} & 0.5\\
$H_\textrm{B}$ & Half-saturation density for vascular plants & \si{g_B m^{-2}}  & 800 \\
\hline
\multicolumn{4}{l} {\textit{Sphagnum} biomass equation}\\
\hline
$r_\textrm{S}$ & \textit{Sphagnum} growth parameter & \si{yr^{-1}} & 0.2 \\
$S_\textrm{max}$ & Maximum density of \textit{Sphagnum} mosses & \si{g_S m^{-2}} & 800 \\
$k_\textrm{SB}$ & Feedback coefficient for \textit{Sphagnum} &  \si{g_S g_B^{-1} yr^{-1}} & 0.1\\
$H_\textrm{S}$ & Half-saturation density for vascular plants & \si{g_S m^{-2}}  & 300\\
$D_\textrm{S}$ & Diffusion coefficient for \textit{Sphagnum} & \si{m^2 yr^{-1}} & 0.2\\
\hline
\multicolumn{4}{l} {Hydraulic head equation}\\
\hline
$p$ & Precipitation & \si{m yr^{-1}} & 0.5$^*$ \\
$t_\textrm{v}$ & Transpiration parameter & \si{m^3 g_B^{-1} yr^{-1}} & 0.005$^*$ \\
$e$ & Evaporation parameter & \si{m yr^{-1}} & 0.3$^*$ \\
$k$ & Hydraulic conductivity & \si{m yr^{-1}} & 500 \\
$\theta$ & Soil porosity & Dimensionless & 0.7 \\
\hline
\multicolumn{4}{l} {Nutrient availability equation}\\
\hline
$N_\textrm{in}$ & Nutrient input & \si{g_N m^{-2} yr^{-1}} & 0--5\\% (Pattern driving parameter)\\
$u$ & Vascular plant uptake parameter & \si{m^3 g_B^{-1} yr^{-1}} & 0.002 \\
$r$ & Nutrient loss parameter & \si{yr^{-1}} & 0.1 \\
$D_\textrm{N}$ & Diffusion coefficient for nutrients & \si{m^2 yr^{-1}}& 10
\end{tabular}
\caption{Parameter and variables symbols, units and default values~\cite{Rietkerk2004,Eppinga2009a}. Superscript $^*$ denotes values affected by climate. \si{g_B}, \si{g_S}, and \si{g_N} respectively denote grams of vascular biomass, \textit{Sphagnum} biomass, and nutrients. }
\label{table:parameters}
\end{table*}

The effect of climatic stress was investigated by plugging the data from the $2\times\textrm{CO}_2$ scenario of the Canadian Climate Center Second generation General Circulation Model (GCMII)~\cite{McFarlane1992}. We selected the simulation results from a peatland-rich area in Siberia (grid point of $\SI{3.75}{\degree} \times \SI{3.75}{\degree}$ size centred on \SI{90}{\degree} East, \SI{61.23}{\degree} North) and we used the yearly averages of the precipitation, as well as the evaporation and transpiration parameters from the GCM as inputs for our simulations. Conversely, the dependence with the climate scenarios of the nutrient input rate $N_\textrm{in}$ was not considered in the present work.

%%%%%%%%%%%%%%%%%%
\section*{Results and discussion}

\subsection*{Pattern analysis}

\begin{figure*}[tb]
\centering
\includegraphics[width=0.8\textwidth]{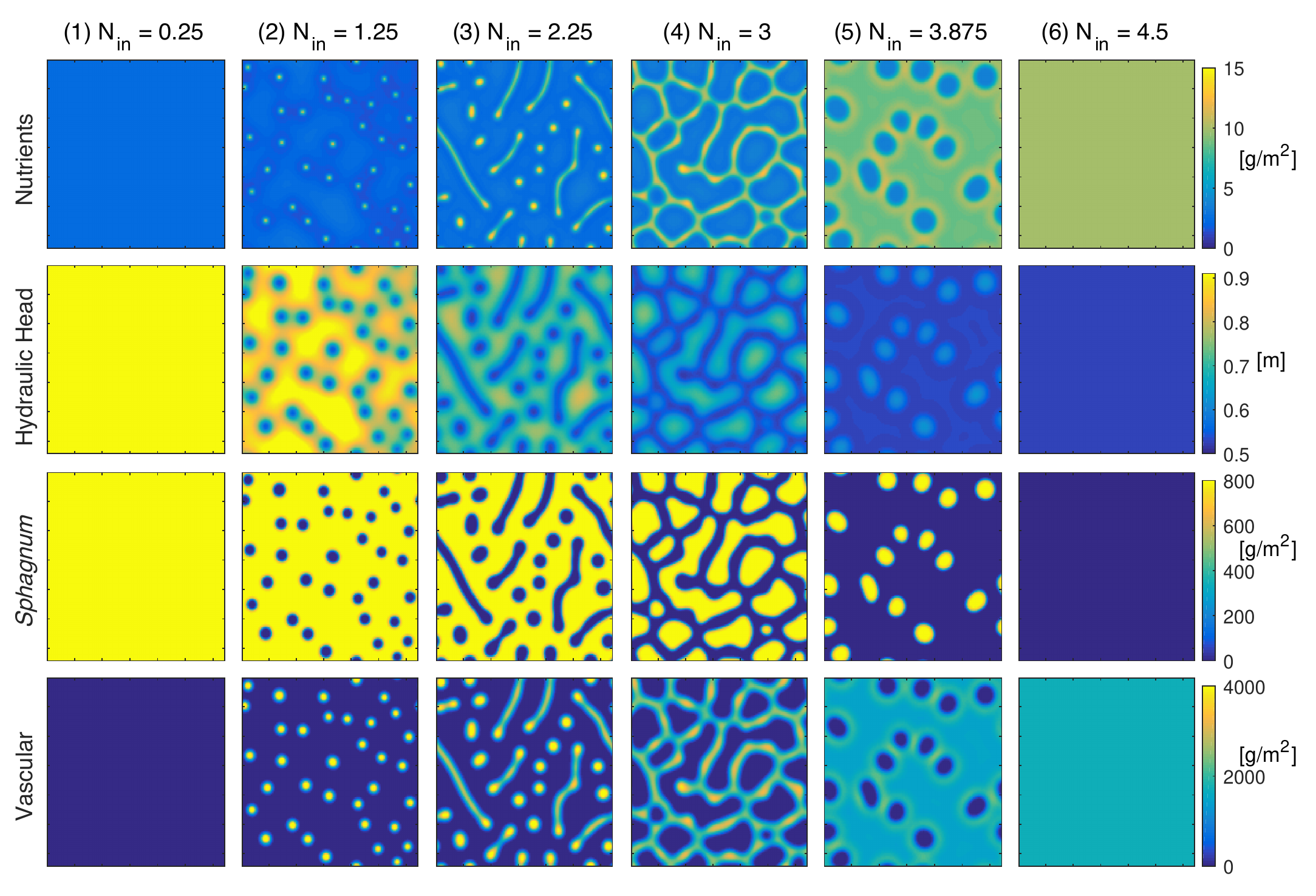}
\caption{Simulated peatland patterns after 400 years stabilization. Initial conditions and parameters are as specified in the Methods section and in \Cref{table:parameters}. From top to bottom: nutrient availability distribution, water level, \textit{Sphagnum} biomass, and vascular plant biomass. Columns 1--6 correspond to nutrient inputs $N_{in}=$ 0.25, 1.25, 2.25, 3, 3.875, and \SI{4.5}{g/m^2/yr}, respectively, as specified above each column. Each panel represents a square 256-meters wide area, with \SI{2}{m} resolution. See Supplementary Movie 1 for the pattern build-up from random initial conditions for $N_\textrm{in}=\SI{2.5}{g/m^2/yr}$.}
\label{fig:patterns}
\end{figure*}

\Cref{fig:patterns} illustrates the variety of peatland patterns in the framework of the nutrient concentration model, as already pointed out by Rietkerk \textit{et al.}~\cite{Rietkerk2004a}. It displays the patterns observed after 400 years, once a stable regime is reached. Initial conditions are described in the Methods below and parameters in \Cref{table:parameters}. Six qualitatively distinct patterns can be distinguished, for increasing external nutrient input $N_\textrm{in}$: (1) full \textit{Sphagnum} coverage; (2) sparse quasi-circular spots of vascular plants; (3) lines and spots of vascular plants, interleaved with a percolating \textit{Sphagnum} network; (4) percolating network of vascular plants; (5) dominance of vascular plants with isolated \textit{Sphagnum} spots; and (6) full vascular plant coverage. Note that, in this work, we define percolation as the existence of at least one cluster continuously connecting two opposite sides of the simulation area.
Although patterns form much earlier in the simulations (See Supplementary Movie 1), they keep evolving for long times and only stabilise after about \SI{400}{years}. Although we do not explore it here, the evolution of peatland over shorter time duration is also of interest (see for example Morris et al.~\cite{Morris2013} for evolution over a decadal range).

{Comparing the patterns for \textit{Sphagnum} and Vascular plant biomasses illustrates the mutual exclusion between the two species, which also explains the sharpness of the patterns. The anti-correlation between the vascular plant biomass and the hydraulic head is due to the water depletion by the vascular plant evapotranspiration. The nutrient availability pattern is slightly more complex. Nutrient maxima at the vascular plant hot spots are surrounded by a nutrient-depleted region, as is especially visible in the second panel of the first row. Due to this structure, vascular plant regions are self-attracting at short distance and self-repulsive at medium range, over a distance governed by the nutrient harvesting efficiency of the vascular plants. This distance governs both the size of the vascular plant clusters (either diameter, or line width), and the distance between them, i.e., the system quasi-periodicity.}
Furthermore, Regimes 1 and 6, 2 and 5, and 3 and 4, respectively, correspond to each other by exchanging the roles of vascular plants and \textit{Sphagnum}. The variety of the patterns is typical of a Turing instability driven by a reaction-diffusion model with mutual retroactions and differential diffusion speeds~\cite{Kukushkin1999,Rietkerk2008,Oppo2009}. Here, the fast diffusion of water and nutrients as compared to that of biomass provides the adequate contrast for generating dots, lines, branched structures and percolating clusters from the same species, depending on a single control parameter, $N_\textrm{in}$.

\begin{figure}[tb]
\centering
\includegraphics[width=0.6\columnwidth]{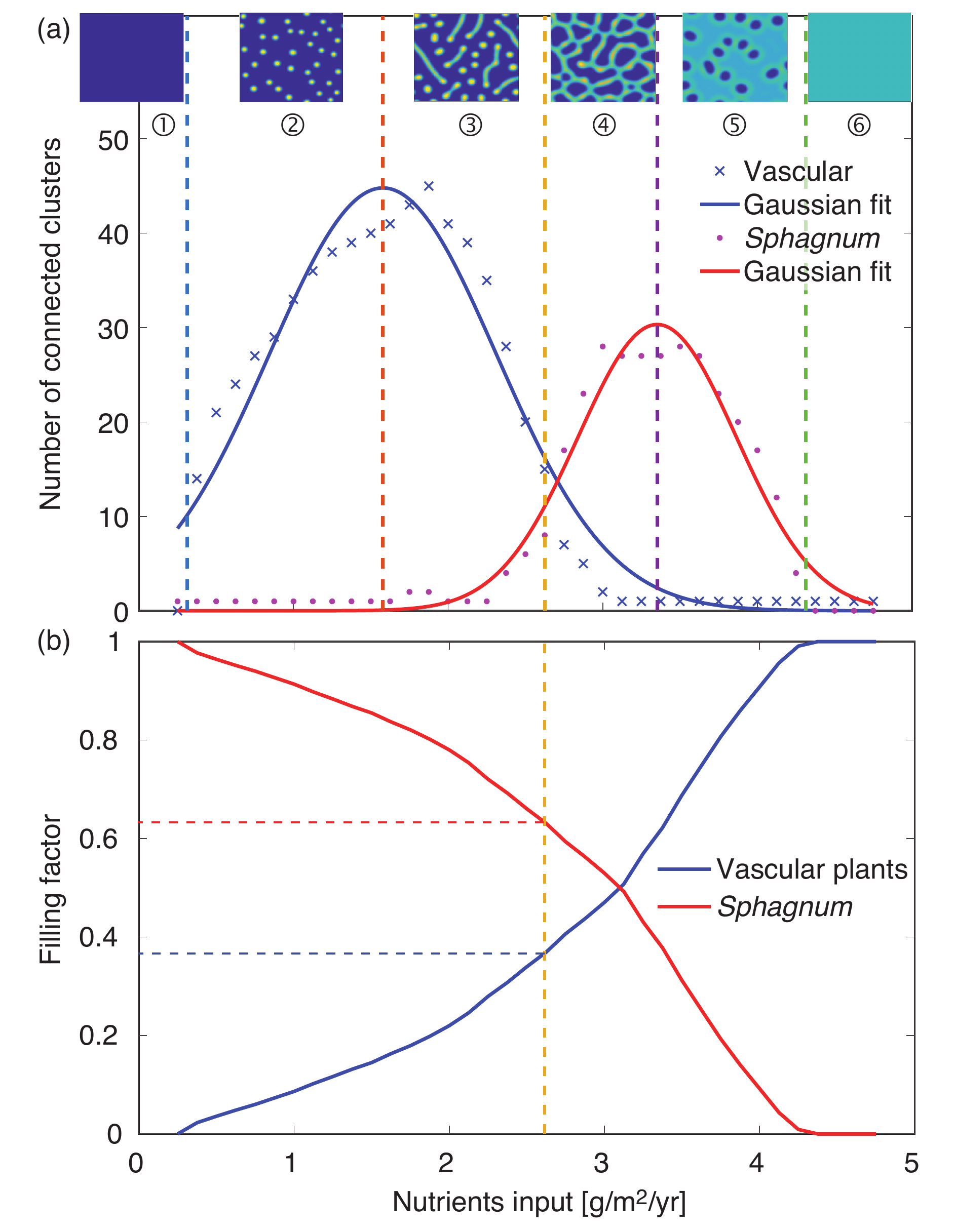}
\caption{(a) Number of clusters as a function of nutrient input. Insets : typical patterns of vascular plant biomass in each range, for $N_{in}=$ 0.25, 1.25, 2.25, 3, 3.875, and \SI{4.5}{g/m^2/yr}. The vertical dotted lines at $N_{in}=$ 0.31, 1.58, 2.63, 3.35, and \SI{4.41}{g/m^2/yr} correspond to the transitions between regimes. (b) Filling factor (with a binarization threshold at 500 and \SI{200}{g/m^2} for vascular plant and \textit{Sphagnum} biomass, respectively) as a function of the nutrient input. 
See \Cref{table:parameters} and Methods for parameter values and initial conditions.}
\label{fig:nb_clusters}
\end{figure}

\begin{figure}[tb]
\centering
\includegraphics[width=0.6\columnwidth]{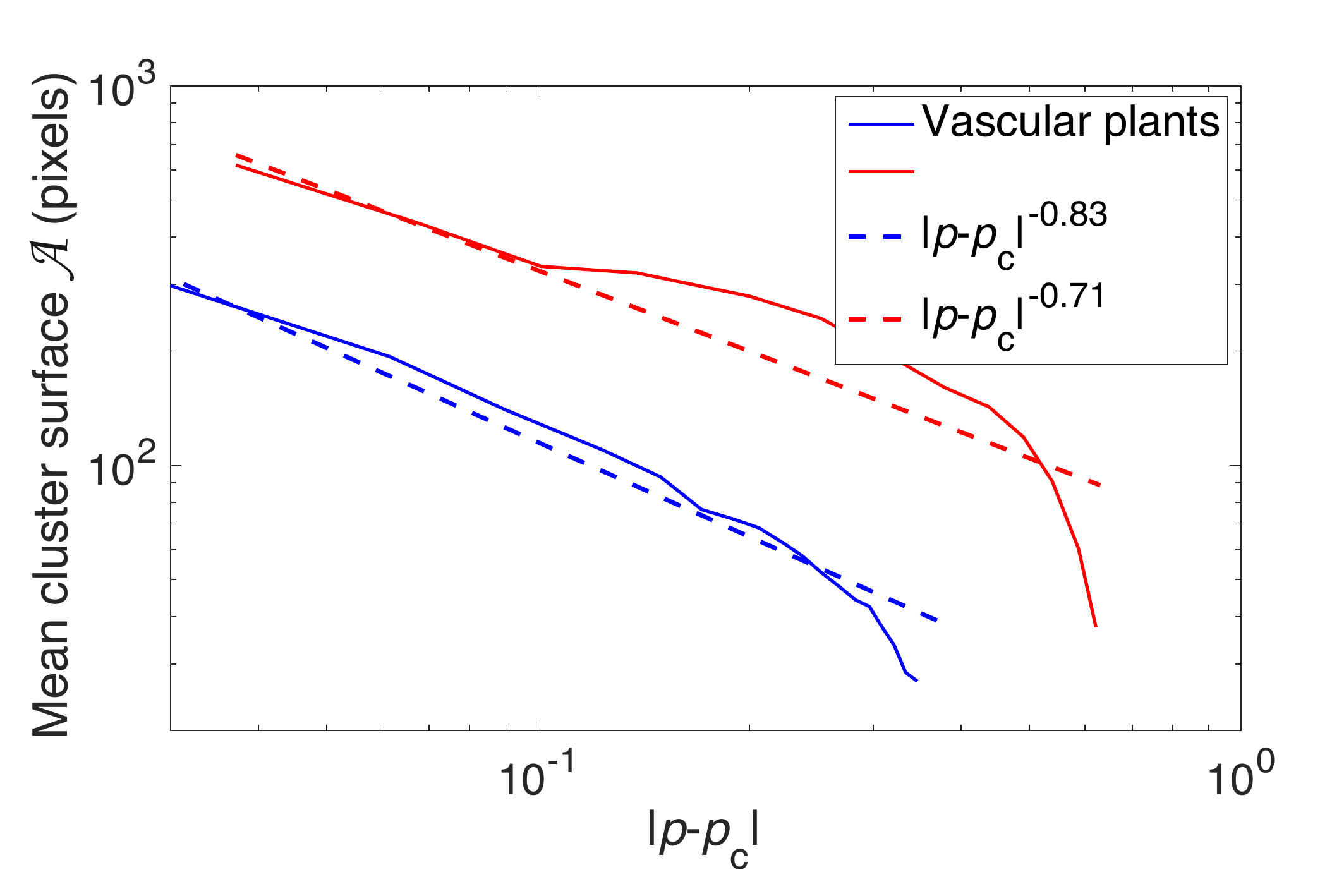}
\caption{Power scaling of the average cluster area $\mathcal{A}$ in the vicinity of the percolation transition. The filling factor is calculated by binarizing the patterns at 500 and \SI{200}{g/m^2} for vascular plant and \textit{Sphagnum} biomass, respectively.}
\label{fig:fit_perco}
\end{figure}

Beyond the qualitative description of the patterns and the understanding of the interactions between the model constituents, our aim is to provide a quantitative characterization and systematic sorting of the clusters of \textit{Sphagnum} and of vascular plants, respectively. To this end, we plotted the number of clusters of both vascular plants and \textit{Sphagnum} as a function of $N_\textrm{in}$ (\Cref{fig:nb_clusters}). In the low-nutrient regime (Regime 1) neither vascular plant clusters, nor non-percolating \textit{Sphagnum} clusters are observed. The jump into Regime 2 corresponds to the point where nutrients are sufficient to allow the long-term survival of sparse vascular plant clusters, the number of which increases with the nutrient input $N_\textrm{in}$.
Simultaneously, their biomass density increases and their nearest-neighbour distance decreases, allowing interactions between neighbouring clusters. They start merging into linear clusters, defining a second threshold above which the number of vascular plant clusters starts decreasing. Progressively, these linear clusters branch and tend to isolate \textit{Sphagnum} regions, allowing the appearance of non-percolating \textit{Sphagnum} clusters in increasing numbers.

The transition from \textit{Sphagnum} percolation to vascular plant percolation defines the transition from Regimes 3 to 4. 
Although the percolation threshold is characterised independently from the cluster number (See Methods), it approximately corresponds to the crossing between the decaying number of vascular plant clusters, and the rising number of \textit{Sphagnum} clusters. 

The initially disconnected vascular plant clusters then progressively merge into the percolating one. Simultaneously, the number of \textit{Sphagnum} clusters increases, as the \textit{Sphagnum} domains divide into smaller entities in reaction to the emerging connexions between the vascular plant clusters. This number reaches a maximum where the \textit{Sphagnum} clusters have their minimum stable size. In Regime 5 this number decreases again because the vascular plants diffuse faster and spread at the expense of \textit{Sphagnum} clusters. When the latter are below their minimum stable size, they finally vanish. Note that during this process, the \textit{Sphagnum} biomass density per unit surface where it is present is stable and close to its maximum value $S_\textrm{max}$, so that the \textit{Sphagnum} decay mainly occurs via the reduction of the surface it covers.
Finally, in Regime 6, \textit{Sphagnum} has fully disappeared under the pressure of the vascular plants. 

We characterized the percolation transition by determining the scaling law and the associated critical exponent near to it. We fitted the mean cluster area $\mathcal{A}$ (excluding the percolating cluster) of both vascular plant and \textit{Sphagnum}, as a power law of the filling factor $p$, i.e., the fraction of the surface covered by the considered species (\Cref{fig:nb_clusters}b) in the vicinity of the transition. This transition occurs at the critical filling factor $p_\textrm{c}$ ($p_\textrm{c,V} = 0.37$ and $p_\textrm{c,S} = 0.63$ for binarization thresholds of 500 and \SI{200}{g/m^2} for vascular and \textit{Sphagnum} biomass, respectively):
\begin{equation}
\mathcal{A} \propto |p-p_\textrm{c}|^{-\gamma}
\end{equation}

We obtain $\gamma_\textrm{V} = 0.83 \pm 0.08$ and $\gamma_\textrm{S}=0.71\pm0.15$ for the vascular plant and Sphanum biomass, respectively (\Cref{fig:fit_perco}). Although the filling factors depend on the binarization threshold, the latter has little influence on the critical exponents within the ranges discussed in the Methods below.
These values are much lower than observed in other experimental percolating systems~\cite{Ettoumi2015} as well as the theoretical prediction of $43/18\approx2.34$ for two-dimensional uncorrelated percolation. This smoother transition to percolation in our model expectedly stems from local correlations due to the plant growth and expansion dynamics, as well as the various feedbacks between the biomass, hydraulic head and nutrient concentration.

\begin{figure}[tb]
\centering
\includegraphics[width=0.5\columnwidth]{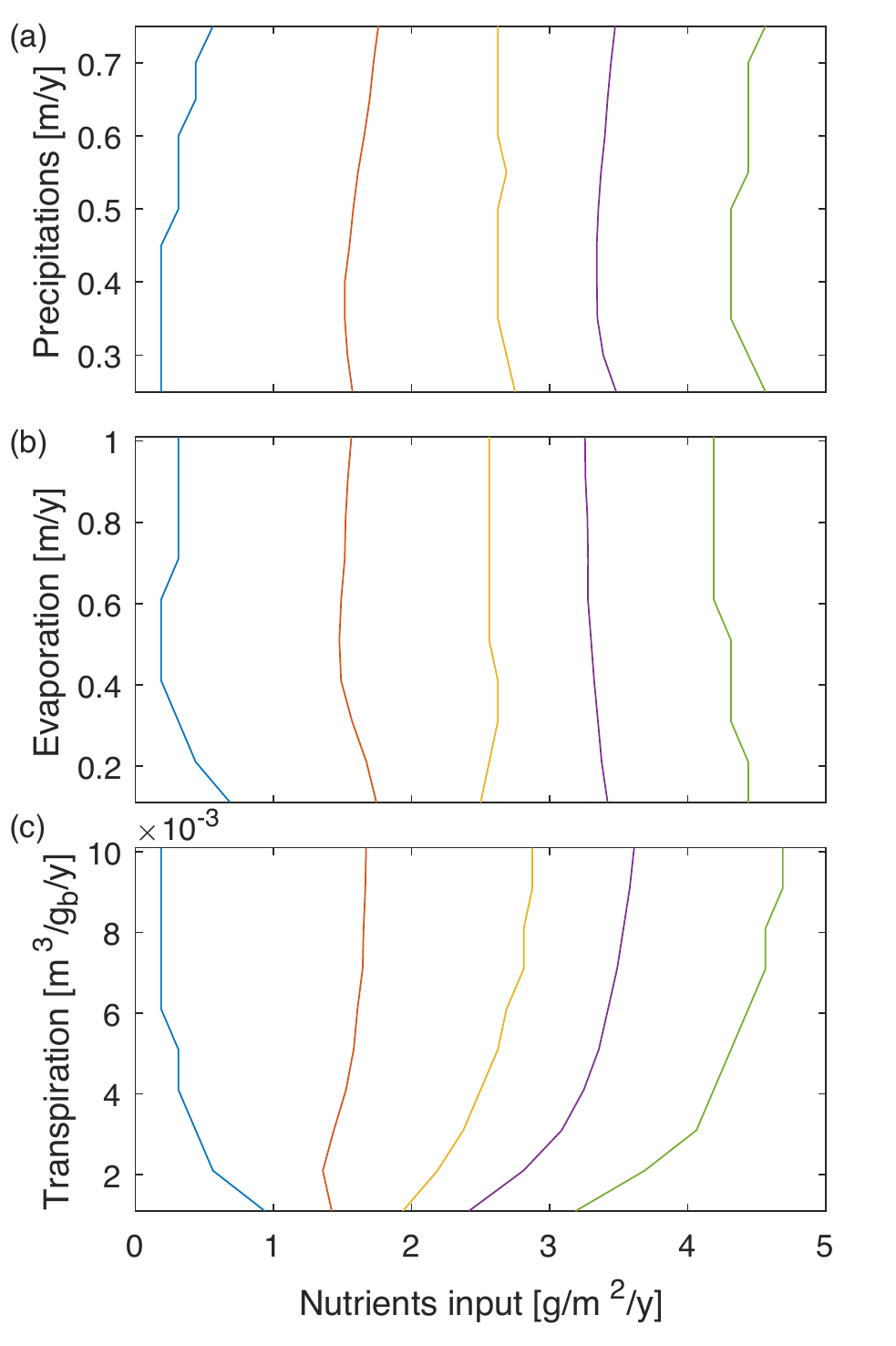}
\caption{Thresholds of the self-patterning regimes as a function of (a) precipitation, (b) evaporation parameter $e$, and (c) transpiration parameter $t_\textrm{v}$. See \Cref{table:parameters} and Methods section for parameter values and initial conditions.}
\label{fig:thresholds}
\end{figure}

\subsection*{Effect of climate on peatland patterning}

The above-described sorting of self-organized patterns and characterization of the transitions between them allows to investigate the pattern changes and therefore their stability, more quantitatively. Here, we illustrate this ability in the case of climate change.
 
The model parameters that may be expected to be sensitive to climate change are the ones related to the hydrological fluxes: precipitations, evaporation, and transpiration. The latter two strongly depend on temperature. For example, the evapotranspiration is usually expressed by the Penman-Monteith equation~\cite{Allen1998}.
Moreover, the evaporation and transpiration terms of \Cref{eq:H} display similar trends when atmospheric conditions change. Experimental studies showed that both are mainly governed by the water availability in the soil, i.e., the water-stress function $f(H)$.
In particular, the transpiration parameter $t_\textrm{v}$ is quite insensitive to changes in temperature: in hot conditions transpiration cooling tends to compensate the reduced evaporation related to photosynthesis~\cite{Kim1996,Drake2018}.
Climate change is therefore implemented into the model by changing the precipitation level as well as by multiplying the evapotranspiration term by a Penman-Monteith factor that is directly provided by the GCMII.

We first performed a single-parameter sensitivity analysis to estimate the model response to a change in the climatic variables.
As displayed in \Cref{fig:thresholds}a,b, the transitions between the six regimes depend little on precipitation and evaporation, within a wide range of realistic values. These parameters therefore have little impact, if any, on the peatland patterns in the framework of the model. Beyond the visual observation of similar patterns, our approaches is more quantitative and supports an interpretation in terms of nutrient accumulation of previous assumptions that peatlands are robust to climate change~\cite{Kettridge2014,Waddington2015,Schneider2016}. 
This stability also illustrates the fact that within our peatland model, changes in net water supply are not critical, while the parameter primarily governing the pattern formation in this system is nutrient availability and its local accumulation by vascular plants.
In contrast, strongly decreasing the vascular plant transpiration parameter tends to restrict the range of nutrient input over which the transitions occur (\Cref{fig:thresholds}c). This can be understood by considering that a reduced vascular plant transpiration will reduce the hydraulic head gradients and the associated nutrient transport via the groundwater flow (see last term in \Cref{eq:N}). At low external nutrient input $N_\textrm{in}$, the reduced nutrient flow towards vascular plants limits their ability to harvest nutrients. Therefore, the threshold for their survival increases. At higher nutrient input, vascular plants can develop, but the less efficient transport of nutrients limits the inhomogeneities in their concentration, hence also in the vascular plant distribution. The latter therefore spread more, so that  the thresholds for their domination over \textit{Sphagnum} is lower.
However, the transpiration parameter $t_\textrm{v}$ depends mostly on the plant species and the water-stress function $f(H)$, rather than on atmospheric temperature and humidity. In particular, Kim and Verma~\cite{Kim1996} showed little deviation between the measured and potential evaporation. Furthermore, Drake \textit{et al.} observed in Eucalyptus that a simulated heat wave induces a simultaneous decrease of photosynthesis and increase of transpiration cooling, with a conserved total water flux, governed by the water availability~\cite{Drake2018}. Therefore, a significant change in evapotranspiration would require a change in the vascular plant species, which is beyond the scope of the present study.

\begin{figure}[t]
	\centering
	\includegraphics[width=0.7\columnwidth]{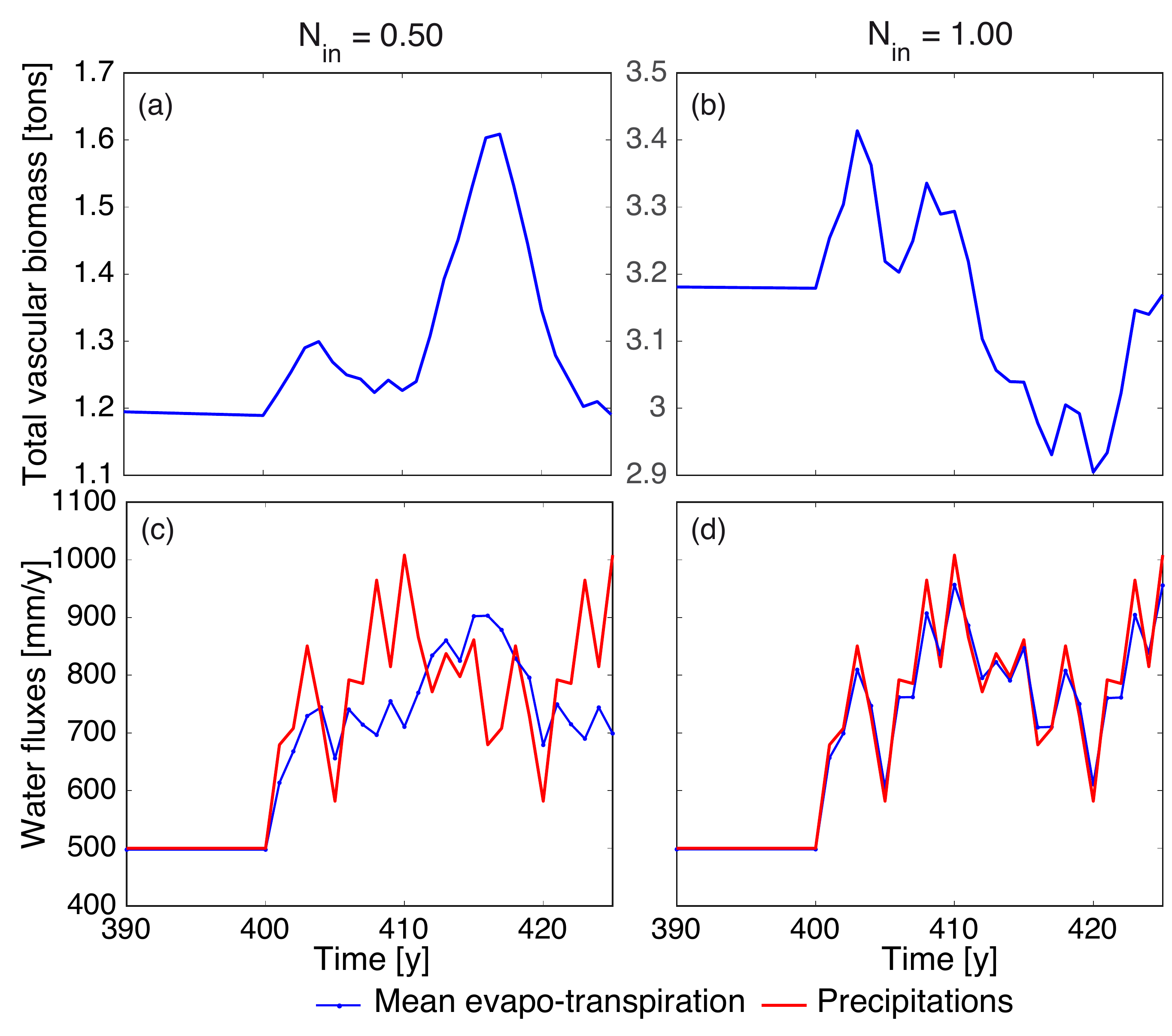}
	\caption{Influence of the $2\times$CO$_2$ scenario of the GCMII on a patterned peatland. (a,b) Evolution of total vascular biomass on simulation area; (c,d) Precipitation and spatially-averaged evapotranspiration. Climate fluctuations are turned on after stabilising the patterns over 400 years.
	See also the corresponding pattern stability in Supplementary Movie 2.}
	\label{fig:climate}
\end{figure}

To further investigate the stability of peatland patterning under climate stress beyond the variation of individual parameters, we simulated a change in the climate. After stabilizing the patterns over 400 years, we switched the precipitation and evapotranspiration rates to follow those from the $2\times\textrm{CO}_2$ scenario of the GCMII.
As expected from the stability of the thresholds against changes in individual parameters, the patterns remain very stable in response to the climate change (See the behavior before year 415 and after year 424, i.e., outside of the drought time, in Supplementary Movie 2). This is observed over a wide range of nutrient input, namely from 0.25 to \SI{4.75}{g m^{-2} yr^{-1}}. The system equilibrates by adapting the vascular plant biomass (\Cref{fig:climate}a,b) to the available water, stabilizing the hydraulic head gradient and level  by increasing the water losses by transpiration to compensate the precipitation level changes (\Cref{fig:climate}c,d). This feedback is faster (at the year-scale) for high nutrient availability ($N_{in} \ge \SI{0.75}{g m^{-2} yr^{-1}}$), and slows down when the input nutrient flux is lower (see \Cref{fig:climate}). There are two thresholds in the response. The first one corresponds to the resilience of vascular biomass (i.e., the transition from Regimes 1 to 2, see \Cref{fig:patterns,fig:nb_clusters}). Without vascular biomass (Regime 1), evapotranspiration corresponds to evaporation alone and the total water flux is not balanced. From Regime 2 on, the system responds (\Cref{fig:climate}a,c) with a certain time-lag. For $N_{in} \geq \SI{0.75}{g m^{-2} yr^{-1}}$, there is enough vascular biomass in the system to respond rapidly (yearly time scale) to variations in precipitation (\Cref{fig:climate}b,d). 
This feedback ensures the resilience of the system. By influencing the density of vascular biomass, it will also expectedly influence the efficiency of carbon storage in the peatland.

\begin{figure}[tb]
\centering
\includegraphics[width=0.5\columnwidth]{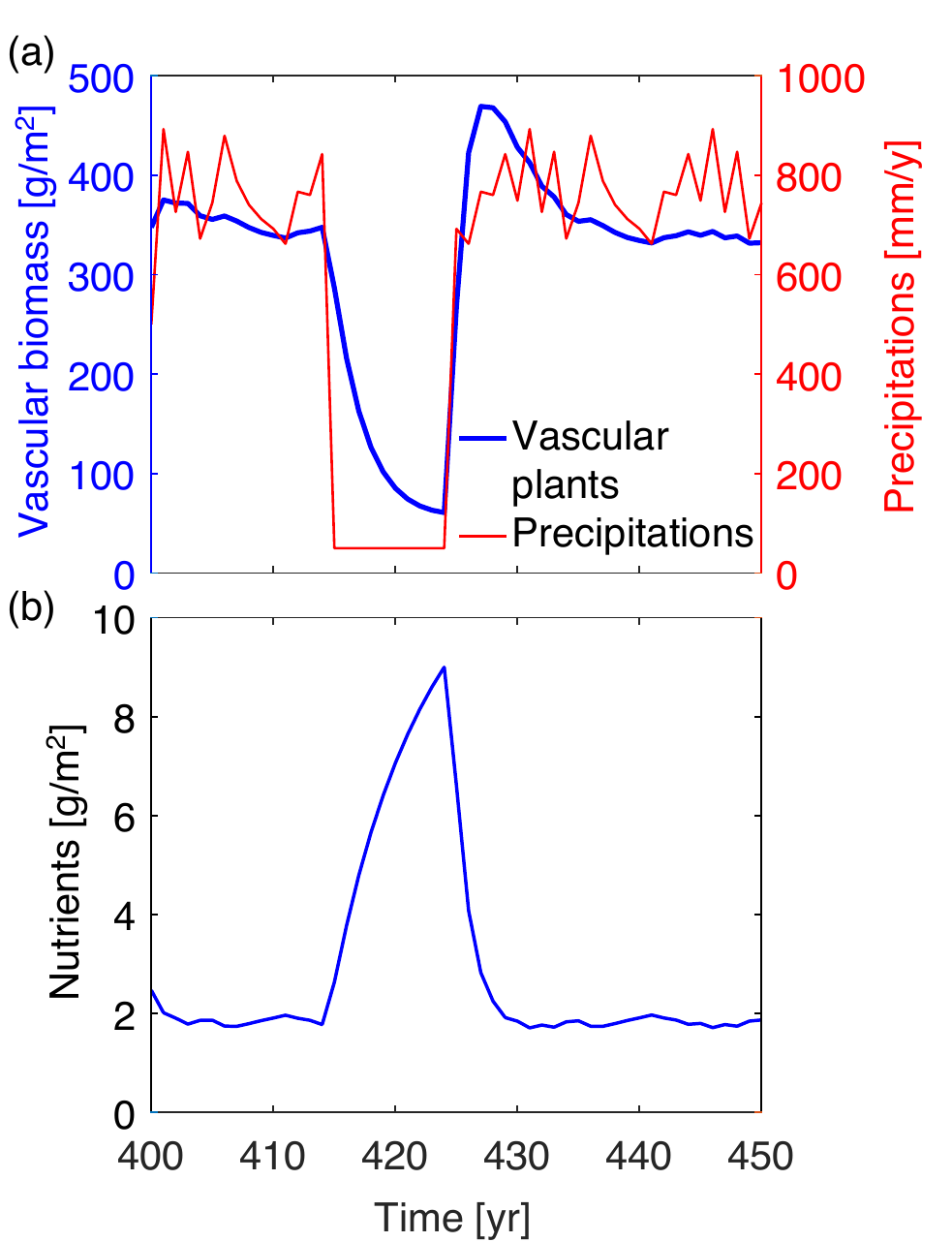}
\caption{Peatland response to a 10-years long drought. (a) Temporal evolution of spatial mean of vascular biomass and precipitations; (b) Spatial mean of nutrient availability. The initial condition before the drought is a peatland stabilized after \SI{400}{years} under the same conditions as in \Cref{fig:patterns}, with $N_\textrm{in} = \SI{1.375}{g m^{-2} yr^{-1}}$, i.e., in Regime 2. Climate conditions are then switched to the $2\times\textrm{CO}_2$ scenario of the GCMII, except for years 415--424 where a drought is simulated by reducing precipitation to \SI{50}{mm/year}.
See also Supplementary Movie 2 illustrating pattern resilience during the drought of years 415--424.}
\label{fig:drought}
\end{figure}

The strength of these feedbacks even allows the patterns predicted by our model to promptly recover after an extreme drought. \Cref{fig:drought} displays the effect on the spatial mean of vascular biomass and nutrients of a 10-years long drought reducing the precipitations from a mean value of \SI{800}{mm/year} to \SI{50}{mm/year}. The pattern structure is not affected by this severe drought (See years 415--424 in Supplementary Movie 2), but the level of biomass strongly decreases, with a drop by 67\% after \SI{10}{years}.
Correspondingly, the nutrient intake by vascular plants is reduced, so that nutrients accumulate in the soil during the drought (\Cref{fig:drought}b). These extra nutrients allow an efficient recovery of the vascular plant biomass at the end of the drought, resulting in a 4-years-long overshoot of up to 35\% above the steady-state vascular biomass density observed before the drought. Finally, once the stored nutrients are consumed, the system recovers to its initial state within 12 years of the end of the drought.

Similar behaviours are observed for various magnitudes and durations of the drought, as well as over a wide range of nutrient inputs (\SIrange{0.25}{4.75}{g m^{-2} yr^{-1}}, corresponding to the Regimes 2--6 displayed in \Cref{fig:patterns,fig:nb_clusters}). The magnitude of the peatland response to a drought depends on the duration and the precipitation level of the drought, as well as on the nutrient input. On the other hand, the recovery time appears to be essentially independent from both the hydrology and nutrient input. 

It should however be kept in mind that the model we use in this work is simplified. In particular, the hydrology is reduced to a homogeneous water table level. Hence, the impact of water shortage is handled via the water-stress function $f$ in the evolution of the vascular plants, but the model does not allow for a full depletion of the water table and therefore does not describe the decay of the \textit{Sphagnum} mosses biomass under dry conditions (see \Cref{eq:S}). Considering \textit{Sphagnum} decay would expectedly result in a decrease of the pattern contrast~\cite{Eppinga2009b}, but without major influence on the rest of the model as the  only coupling considered here is via the competition for space with vascular plants. Due to these limitations, the model outcome as described above cannot be expected to provide quantitative predictions, but rather indications about trends like the robustness of the nutrient accumulation mechanism, and a testbed for interpreting empirical observations. Biologically and ecologically realistic predictions would require a much more detailed and complex model, as discussed above.

\section*{Conclusion}

As a conclusion, based on a diffusion-reaction model of an ombrotrophic peatland, we have proposed a detailed analysis of vegetation pattern, with a quantitative description of the pattern transitions and a systematic sorting of the self-organized patterns. We characterized the transition from \textit{Sphagnum} to vascular plant cluster percolation when the nutrient availability is increased, as well as the associated scaling laws close to the transition.
Such classification and characterization is not specific to the model of Eppinga \textit{et al.}~\cite{Eppinga2009a}, but can straightforwardly be extended to any system displaying Turing instabilities, e.g., reaction-diffusion models with differential spatial mobilities. This covers many ecosystems~\cite{Rietkerk2004b,Rietkerk2008}, including non-ombrotrophic peatlands~\cite{Ogden2005}, mussel banks~\cite{Koppel2008}, coral reefs~\cite{Mistr2003}, drylands~\cite{Klausmeier1999}, or savanna~\cite{Callaway2002}. 
 
Applying our pattern classification to the self-organized patterns generated by a peatland model based on nutrient accumulation against climate stress, including against strong prolonged drought, we evidenced his high resilience. This finding may help understand the empirically observed high peatland stability and resilience~\cite{Kettridge2014,Waddington2015,Schneider2016,Thompson2017}.

%%%%%%%%%%%%%%%%%%
\section*{Methods}

\subsection*{Numerical implementation}

Simulations were performed in Matlab on a 128 $\times$ 128 pixel grid with periodic (toroidal) boundary conditions, covering 256 $\times$ \SI{256}{m}, using a Newton (Euler) numerical scheme with a temporal step d$t$=\SI{e-4}{year}. 
We started from an initial random distribution of spots bearing \SI{200}{g m^{-2}} biomass of either vascular plants or \textit{Sphagnum} mosses, each covering \SI{12.5}{\%} of the grid elements. We checked that the results are independent from the particular initial random pattern. The hydraulic head and nutrients are  homogeneously set to their plantless equilibrium level $H_0 = p/e$ and $N_0 = N_\textrm{in}/r$, respectively~\cite{Rietkerk2004a}.
The system required about \SI{150}{yr} to reach a quasi-steady-state pattern, as displayed in Supplementary Movie 1.

\subsection*{Pattern analysis}

The pattern analysis was performed on vascular plant and \textit{Sphagnum} biomass data, to infer "ridge" and "hollow" components of the landscape. The patterns were binarized, with a threshold at 500 and \SI{200}{g m^{-2}} for vascular plants and \textit{Sphagnum}, respectively. The patterns, the nutrient inputs ensuring the transitions between pattern types, and the critical exponents close to the percolation transition are quite insensitive to these thresholds, over a wide range of values (100 -- \SI{1400}{g m^{-2}} and 50 -- \SI{600}{g m^{-2}}, respectively for vascular plants and \textit{Sphagnum}). In contrast, the absolute values of the filling factors varied with these thresholds. We then detected each cluster of the considered species as a connected component using the Matlab function \textit{bwconncomp} and determined their characteristics (size, surface, shape, etc.) using the Matlab function \textit{regionprops}.  
The number of clusters of both vascular plants and \textit{Sphagnum}, displayed as a function of nutrient input $N_\textrm{in}$, was fitted with Gaussian functions. As detailed in the Results section above, the peak of each Gaussian function defines the transition from approximately circular to elongated clusters.

We defined percolating clusters as clusters continuously connecting two opposite sides of the simulation grid. Due to the periodic boundary conditions, this definition does not exactly match the standard one: percolating clusters cover a full turn around either the small or the large diameter of the torus defined by the periodic boundary conditions. The jump from percolating \textit{Sphagnum} to percolating vascular plants defines a threshold in $N_\textrm{in}$. 
As for other experimental systems like filamentation patterns in high-power laser beams~\cite{Ettoumi2015} or liquid crystals~\cite{Takeuchi2009}, the approach of the percolation point was characterized with a critical exponent $\gamma$, determined by fitting the average cluster size (excluding the percolating cluster) against the filling factor (i.e., the fraction of the surface covered by the considered species) with a power law~\cite{Takeuchi2009}.

%%%%%%%%%%%%
% biblio

%\newpage

%\bibliographystyle{apalike}
\bibliography{Biblio}

\section*{Acknowledgements}

This work was supported by the FNS (grants 200021\_155970 and 200020\_175697).
We gratefully acknowledge fruitful discussions with S.~Goyette, G. Rohat, and technical support by M. Moret. Part of the computations were performed at University of Geneva on the Baobab cluster.

\section*{Author contributions statement}

JK designed the study; CB developed the model and performed the simulations; CB, MB and JK analyzed the results; All authors reviewed the manuscript. 

\section*{Competing Interests}
The authors declare no competing interests

\section*{Data availability statement}
The datasets generated during and/or analysed during the current study are available from the corresponding author on reasonable request.

\newpage

%%%%%%%%%%%%%%%%%
\section*{Supplementary Materials}

\ 

Supplementary Movie 1: Temporal evolution of vascular and \textit{Sphagnum} mosses biomass, nutrient availability and hydraulic head over 400 years in an ombrotrophic peatland. 
 Initial conditions and parameters as specified in the Methods section and in \Cref{table:parameters}. Nutrient input $N_{in}=$\SI{2.5}{g/m^2/yr}. Each panel represents a square 256-meters wide area, with \SI{2}{m} resolution.

 \ 
 
 Supplementary Movie 2: Temporal evolution of vascular plant patterns in an ombrotrophic peatland area of 256 by 256 m, after the switch to the $2 \times$ CO$_2$ scenario of GCMII, for the values of $N_\textrm{in}$ corresponding to the different patterns of \Cref{fig:nb_clusters}: $N_\textrm{in}$ = 0.25, 1.25, 2.25, 3, 3.875, and \SI{4.5}{g/m^2/yr}. Years 415--424 correspond to a drought, with a drop in precipitation to \SI{50}{mm/yr}.

\end{document}